\documentclass{article}
\usepackage{benelearn2014}
\usepackage{graphicx} 
\usepackage{subfigure}
\usepackage[english]{babel}
\usepackage[utf8]{inputenc}
\usepackage{microtype}
\usepackage{amsmath}
\usepackage{amsfonts}
\usepackage{dsfont}

\usepackage{mlapa}

\usepackage{todonotes}
\usepackage{url}

\begin{document}

\twocolumn[
\benelearntitle{Reducing Offline Evaluation Bias in Recommendation Systems}

 \benelearnauthor{Arnaud de Myttenaere}{ademyttenaere@viadeoteam.com}
 \benelearnauthor{Boris Golden}{bgolden@viadeoteam.com}
 \benelearnaddress{Viadeo, 30 rue de la Victoire, 75009 Paris, France}
 \benelearnauthor{Bénédicte Le Grand}{benedicte.le-grand@univ-paris1.fr}
 \benelearnaddress{Centre de Recherche en Informatique, Université Paris 1
   Panthéon – Sorbonne, 90 rue de Tolbiac, 75013 Paris, France}
 \benelearnauthor{Fabrice Rossi}{Fabrice.Rossi@univ-paris1.fr}
 \benelearnaddress{SAMM EA 4534,  Université Paris 1
   Panthéon – Sorbonne, 90 rue de Tolbiac, 75013 Paris, France}
\benelearnaddress{{\bf Keywords}: recommendation systems, offline evaluation,
  evaluation bias, covariate shift}
\vskip 0.3in
]

\begin{abstract}
Recommendation systems have been integrated into the majority of large online
systems. They tailor those systems to individual users by filtering and
ranking information according to user profiles. This adaptation process
influences the way users interact with the system and, as a consequence,
increases the difficulty of evaluating a recommendation algorithm with
historical data (via \emph{offline evaluation}). This paper analyses this
evaluation bias and proposes a simple item weighting solution that reduces its
impact. The efficiency of the proposed solution is evaluated on real world
data extracted from Viadeo professional social network. 
\end{abstract}

\section{Introduction}
A recommender system provides a user with a set of possibly ranked items that
are supposed to match the interests of the user at a given moment
\cite{park2012literature,kantor2011recommender,adomavicius2005toward}. Such
systems are ubiquitous in the daily experience of users of online systems. For
instance, they are a crucial part of e-commerce where they help consumers
select movies, books, music, etc. that match their tastes. They also
provide an important source of revenues, e.g. via targeted ad placements
where the ads displayed on a website are chosen according to the user profile
as inferred by her browsing history for instance. Commercial aspects set
aside, recommender systems can be seen as a way to select and sort information
in a personalised way, and as a consequence to adapt a system to a user.

Obviously, recommendation algorithms must be evaluated before and during their
active use in order to ensure the quality of the recommendations. Live
monitoring is generally achieved using online quality metrics such as the
click-through rate of displayed ads. This article focuses on the offline
evaluation part which is done using historical data (which can be recorded
during online monitoring). One of the main strategy of offline evaluation
consists in simulating a recommendation by removing a confirmation action (click,
purchase, etc.)  from a user profile and testing whether the item associated
to this action would have been recommended based on the rest of the profile
\cite{shani2011evaluating}. Numerous variations of this general scheme are
used ranging from removing several confirmations to taking into account item
ratings.

While this general scheme is completely valid from a statistical point of
view, it ignores various factors that have influenced historical data as the recommendation algorithms previously used.

Assume for instance that several recommendation algorithms are evaluated at
time $t_0$ based on historical data of the user database until $t_0$. Then the best
algorithm is selected according to a quality metric associated to the offline
procedure and put in production. It starts recommending items to the
users. Provided the algorithm is good enough, it generates some confirmation
actions. Those actions can be attributed to a good user modeling but also to
luck and to a natural attraction of some users to new things. This is
especially true when the cost of confirming/accepting a recommendation is
low. In the end, the state of the system at time $t_1>t_0$ has been influenced by the recommendation algorithm in production. 
 
 
 Then if one wants to monitor the performance of this
algorithm at time $t_1$, the offline procedure sometimes overestimates the
quality of the algorithm because confirmation actions are now frequently
triggered by the recommendations, leading to a very high predictability of the
corresponding items.

This bias in offline evaluation with online systems can also be caused by other
events such as a promotional offer on some specific products between a first
offline evaluation and a second one. Its main effect is to favor algorithms
that tend to recommend items that have been favored between $t_0$ and $t_1$
and thus to favor a kind of ``winner take all'' situation in which the
algorithm considered as the best at $t_0$ will probably remain the best one
afterwards, even if an unbiased procedure could demote it. While limits of
evaluation strategies for recommendation algorithms have been identified in
e.g. \cite{HerlockerEtAl2004Evaluating,mcnee2006being,said2013user}, the
evaluation bias described above has not been addressed in the literature, to
our knowledge. 

This paper proposes a modification of the classical offline evaluation
procedure that reduces the impact of this bias. Following the
general principle of weighting instances used in the context of covariate
shift \cite{sugiyama2007covariate}, we propose to assign a tunable weight to each item. The weights are optimized in order to reduce the bias without discarding new data generated since the reference
evaluation. 

The rest of the paper is organized as follows. Section
\ref{sec:problem-formulation} describes in detail the setting and the problem
addressed in this paper. Section \ref{sec:reduc-eval-bias} introduces the
weighting scheme proposed to reduce the evaluation bias. Section
\ref{sec:exper-eval} demonstrates the practical relevance of the method on
real world data extracted from the Viadeo professional social
network\footnote{Viadeo is the world's second largest professional social
  network with 55 million members in August 2013. See
  \url{http://corporate.viadeo.com/en/} for more information
  about Viadeo.}. 

\section{Problem formulation}\label{sec:problem-formulation}

\subsection{Notations and setting}
We denote $U$ the set of users, $I$ the set of items and $\mathcal{D}_t$ the historical data available at time $t$. A recommendation algorithm is a function $g$ from $U\times \mathcal{D}_t$ to some set built from $I$. We will denote $g_t(u) = g(u,\mathcal{D}_t)$ the recommendation computed by $g$ at instant $t$ for user $u$.
The recommendation strategy, $g_t(u)$, could be a list of $k$ items (ordered in
decreasing interest), a set of $k$ items (with no ranking), a mapping from a
subset of $I$ to numerical grades for some items, etc. The specifics are not
relevant to the present analysis as we assume given a quality function $l$ from
product of the result space of $g$ and $I$ to $\mathbb{R}^+$ that measures to
what extent an item $i$ is correctly recommended by $g$ at time $t$ via $l(g_t(u),i)$.

Offline evaluation is based on the possibility of ``removing'' any item $i$
from a user profile ($I_u$ denotes the items associated to $u$). The result is
denoted $u_{-i}$ and $g_t(u_{-i})$ is the recommendation obtained at instant $t$ when $i$ has
been removed from the profile of user $u$. If $g$ outputs a subset of $I$,
then one possible choice for $l$ is $l(g_t(u_{-i}),i)=1$ when $i\in g_t(u_{-i})$
and 0 otherwise. If $g$ outputs a list of the best $k$ items, then $l$
will decrease with the rank of $i$ in this list (it could be, e.g., the
inverse of the rank). 

Finally, offline evaluation follows a general scheme in which a user is chosen
according to some prior probability on users $P(u)$ (these probabilities might
reflect the business importance of the users, for instance). Given a user, an
item is chosen among the items associated to its profile, according to some
conditional probability on items $P(i|u)$. When an item $i$ is not associated
to a user $u$ (that is $i\not\in I_u$), $P(i|u)=0$. Notice than while we use a
stochastic framework, exhaustive approaches are common in medium size
systems. In this case, the probabilities will be interpreted as weights and
all the pairs $(u,i)$ (where $i\in I_u$) will be used in the evaluation
process. In both stochastic and exhaustive evaluations, a very common choice
for $P(u)$ is the uniform probability on $U$. It is also quite common to use a uniform probability for $P(i|u)$. For instance, one could favor items recently
associated to a profile over older ones.

The two distributions $P(u)$ and $P(i|u)$ lead to a joint distribution
$P(u,i)=P(i|u)P(u)$ on $U\times I$. In an online system, $P(i|u)$ evolves over time\footnote{While $P(u)$ could also
  evolve over time, we do not consider the effects of such evolution in
  the present article.}. For example, if the probability $P(i|u)$ is uniform
over the items associated to user $u$, then as soon as $u$ gets a new item
(recommended by an algorithm, for instance), all probabilities are
modified. The same is true for more complex schemes that take into account the
age of the items, for instance. 

\subsection{Origin of the bias in offline evaluation}
The offline evaluation procedure consists
in calculating the quality of the recommender $g$ at instant $t$ as
$L_t(g)=\mathbb{E}(l(g_t(u_{-i}),i))$ where the expectation is taken with respect
to the joint distribution, that is
\begin{equation}\label{eq:loss}
L_t(g)=\sum_{(u,i)\in U\times I}P_t(i|u)P_t(u)l(g_t(u_{-i}),i).
\end{equation}
In very large systems, $L_t(g)$ is approximated by actually sampling from
$U\times I$ according to the probabilities while in small ones, the
probabilities are used as weights, as pointed out above.

 Then if two algorithms are evaluated at two
different moments, their qualities are not directly comparable. While this
problem does not fall exactly into the covariate shift paradigm
\cite{Shimodaira2000227}, it is related: once a recommendation algorithm is chosen based on
a given state of the system, it is almost guaranteed to influence the state of
the system while put in production, inducing an increasing distance between its
evaluation environment (i.e. the initial state of the system) and the
evolving state of the system. This influence of the recommendation algorithm on the state of the system is responsible for the bias since offline evaluation relies on historical data.

A naive solution to this bias would be to define a fixed evaluation database (a snapshot of the
system at $t_0$) and to compare algorithms only with respect to the original
database. This is clearly unacceptable for an online system as it would discard
both new users and, more importantly, evolutions of user profiles. 

\subsection{Real world illustration of the bias}\label{sec:real-world-illustr}
\begin{figure}[htb]
  \centering
\includegraphics[width=\linewidth]{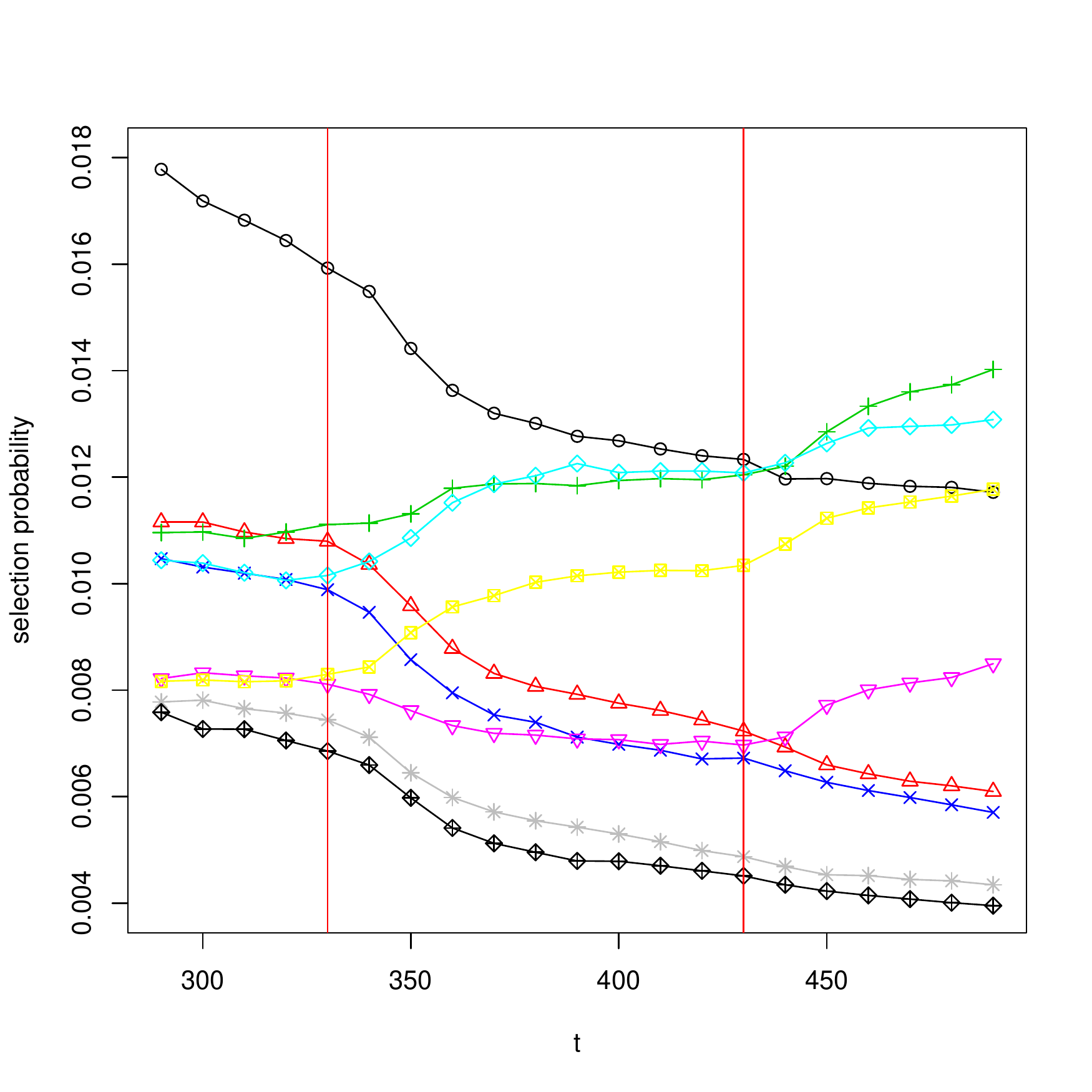}
  \caption{Impact of recommendation campaigns on the item probabilities: each
    curve displays the evolution of $P(i)$ over time for a given item.}
  \label{fig:impact}
\end{figure}

We illustrate the evolution of the $P_t(i)$ probabilities in an online system with
a functionality provided by the Viadeo platform: each user can claim to have
some skills that are displayed on his/her profile (examples of skills include
project management, marketing, etc.). In order to obtain more complete profiles,
skills are recommended to the users via a recommendation algorithm, a practice
that has obviously consequences on the probabilities $P_t(i)$, as illustrated on Figure
\ref{fig:impact}. 

The skill functionality has been implemented at time $t=0$. After 300 days,
some of the $P_t(i)$ are roughly static. Probabilities of other items still
evolve over time under various influences, but the major sources of evolution
are recommendation campaigns.  Indeed, at times $t=330$ and $t=430$,
recommendation campaigns have been conducted: users have received personalized
recommendation of skills to add to their profiles. The figure shows strong
modifications of the $P_t(i)$ quickly after each campaign.  In particular, the
probabilities of the items which have been recommended increase significantly;
this is the case for the black, pink and light blue curves at $t=330$. On the
other hand, the probabilities of the items which have not been recommended
decrease at the same time. The probabilities tend to become stable again until
the same phenomenon can be observed right after the second recommendation
campaign at $t=430$: the curves corresponding to the items that have been
recommended again keep increasing. The green curve represents the probability
of an item which has been recommended only during the second recommendation
campaign. Section \ref{sec:bias-action} demonstrates the effects of this
evolution on algorithm evaluations.

\section{Reducing the evaluation bias}\label{sec:reduc-eval-bias}

\subsection{Principle for reducing the bias}\label{sec:analys-naive-recomm}
Let us consider a naive algorithm which always recommends the same items whatever the user and historical data. In other words, $g$ is constant. Constant algorithms are particularly easy to understand and useful to illustrate the bias due to external factors. Indeed one can reasonably assume that the score of such algorithms does not strongly vary over time.

A simple transformation of equation (\ref{eq:loss}) shows that for a constant algorithm $g$:
\begin{align}
  \label{eq:loss:constant}
L_t(g)&=  \sum_{u\in U}\sum_{i\in I}P_t(i|u)P_t(u)l(g_t,i) \notag\\
&=\sum_{i\in I}l(g_t,i)\sum_{u\in U}P_t(i|u)P_t(u) \notag\\
&=\sum_{i\in I}P_t(i)l(g_t,i).
\end{align}

As a consequence, a way to guarantee a stationary evaluation framework for a
constant algorithm is to have constant values for the $P_t(i)$ (the marginal
distribution of the items). 

A natural solution to have constant values for $P_t(i)$ would be to record those probabilities at $t_0$ and use
them subsequently in offline evaluation as the probability to select an item. However, this would require to revert the way offline evaluation is done: first select an item, then select a user having this item with a certain probability $\pi_t(u|i)$.
But as the probability law originally defined on users reflects their relative importance and should not be modified, it will be necessary to compute $\pi_t(u|i)$ such as the overall probability law on users is close enough to the original one $\big(P_{t_0}(u)\big)$. The computation of the coefficients $\pi_t(u|i)$ would need to be done for all users. Keeping the standard offline evaluation procedure and computing coefficients to alter the probabilities of selecting an item for a given user is more efficient because it can be done only for a limited number of key items (in practice in much smaller quantity than the number of users for most of real world systems) leading to a much lower complexity.


A strong assumption we make is that in practice reducing offline evaluation bias for constant algorithms contributes to reducing offline evaluation bias for all algorithms.

\subsection{Item weights}
$P_t(i|u)$ probabilities are thus the only quantities that can be modified in order to reduce
the bias of offline evaluation. In particular, $P_t(u)$ is driven by
business considerations related to the importance of individual users and can
seldom be manipulated without impairing the associated business
metrics. We propose therefore to depart from the classical values for $P_t(i|u)$
(such as using a uniform probability) in order to mimic static values for
$P_ {t_0}(i)$. This approach is related to the weighting strategy used in the case of
covariate shift \cite{sugiyama2007covariate}.

This is implemented via tunable item specific weights, the $\omega=(\omega_i)_{i\in
  I}$, which induce modified conditional probabilities $P_t(i|u,\omega)$. The
general idea is to increase the probability of selecting $i$ if $\omega_i$ is
larger than 1 and vice versa, so that $\omega$ recalibrates the probability of selecting each item. The simplest way to implement this probability
modification is to define $P_t(i|u,\omega)$ as follows:
\begin{equation}
  \label{eq:weighted:conditional}
P_t(i|u,\omega)=\frac{\omega_iP_t(i|u)}{\sum_{j\in I_t}\omega_jP_t(j|u)}.
\end{equation}

Other weighting schemes could be used. Notice that these weighted conditional
probabilities lead to weighted item probabilities defined by:
\begin{equation}
  \label{eq:weighted:item}
P_t(i|\omega)=  \sum_{u\in  U}P_t(i|u,\omega)P_t(u).
\end{equation}

\subsection{Adjusting the weights}

We thus reduce the evaluation bias by leveraging the weights $\omega$ and using the
associated distribution $P_{t_1}(i|u,\omega)$ instead of
$P_{t_1}(i|u)$. Indeed one can chose $\omega$ in such as way that
$P_{t_1}(i|\omega)\simeq P_{t_0}(i)$. This allows one to use 
all the data available at time $t_1$ for the offline evaluation while limiting the bias induced by those new data.

This leads to a non-linear system with $n_i$ equations and $n_i$ parameters ($\omega_1, \dots, \omega_{n_i}$) such that for all $i \in I_{t_1}$:
\[ \sum_{u \in U} \frac{\omega_i P_{t_1}(i|u,\omega_i)}{\sum_{j \in I}\omega_j P_{t_1}(j|u,\omega_j)}\cdot P_{t_1}(u) =  P_{t_0}(u)\]
$\omega$ cannot be solved easily and we thus need to approximate it using an optimisation algorithm.

Optimizing the weights amounts to reducing a dissimilarity between the
weighted distribution and the original one. We use here the Kullback-Leibler
divergence, that is
\begin{align}
  \label{eq:divergence}\notag
D(\omega)&=D_{KL}(P_{t_0}(.)\|P_{t_1}(.|\omega))\\
&=\sum_{i\in I_{t_0}}P_{t_0}(i)\log\frac{P_{t_0}(i)}{P_{t_1}(i|\omega)}.
\end{align}
Where $I_{t_0}$ represents the set of items which have been selected at least once at $t_0$.

The asymmetric nature of $D_{KL}$ is useful in our context as it reduces
the influence of rare items at time $t_0$ as they were not very important in
the calculation of $L_{t_0}(g)$. 


The target probability $P_{t_0}(i)$ is computed once and for all items at the
initial evaluation time. One coordinate of the gradient can be computed in $\mathcal{O}(N_{U\times I_{t_1}})$, where $N_{U\times I_{t_1}}$ is the number of couples $(u,i)$ with $i \in I_u$ and $u \in U$ at instant $t_1$. Thus the whole gradient can be computed in complexity $\mathcal{O}(n_i \cdot N_{U\times I_{t_1}})$. This would be prohibitive on a large
system. To limit the optimization cost, we focus on the largest
modifications between $P_{t_0}(i)$ and $P_{t_1}(i)$. More precisely, we
compute once $P_{t_1}(i)$ for all $i\in I_{t_0}$ and select the subset $I_{t_0}^{t_1}(p)$ of $I_{t_1}$
of size $p$ which exhibits the largest differences in absolute values between $P_{t_0}(i)$ and $P_{t_1}(i)$.

Then $D(\omega)$ is only optimized with respect
to the corresponding weights $(\omega_i)_{i\in I_{t_0}^{t_1}(p)}$, leading to a cost in
$\mathcal{O}(p\cdot N_{U\times I_{t_1}})$ for each gradient calculation. Notice that $p$ is therefore an
important parameter of the weighting strategy. In practice, we optimize the
divergence via a basic gradient descent.

Notice that to implement weight optimization, one needs to compute
$P_{t_0}(i)$ and $P_{t_1}(i)$. As pointed out in \ref{sec:analys-naive-recomm}
these are costly operations. We assume however that evaluating several
recommendation algorithms has a much larger cost, because of the repeated
evaluation of $L_t(g)$ associated to e.g. statistical model parameter
tuning. Then while optimizing $\omega$ is costly, it allows one to rely on the
efficient classical offline strategy to evaluate recommendation algorithms
with a reduced bias. 

\section{Experimental evaluation}\label{sec:exper-eval}

\subsection{Data and metrics}
\begin{figure}[htb]
  \centering
\includegraphics[width=\linewidth]{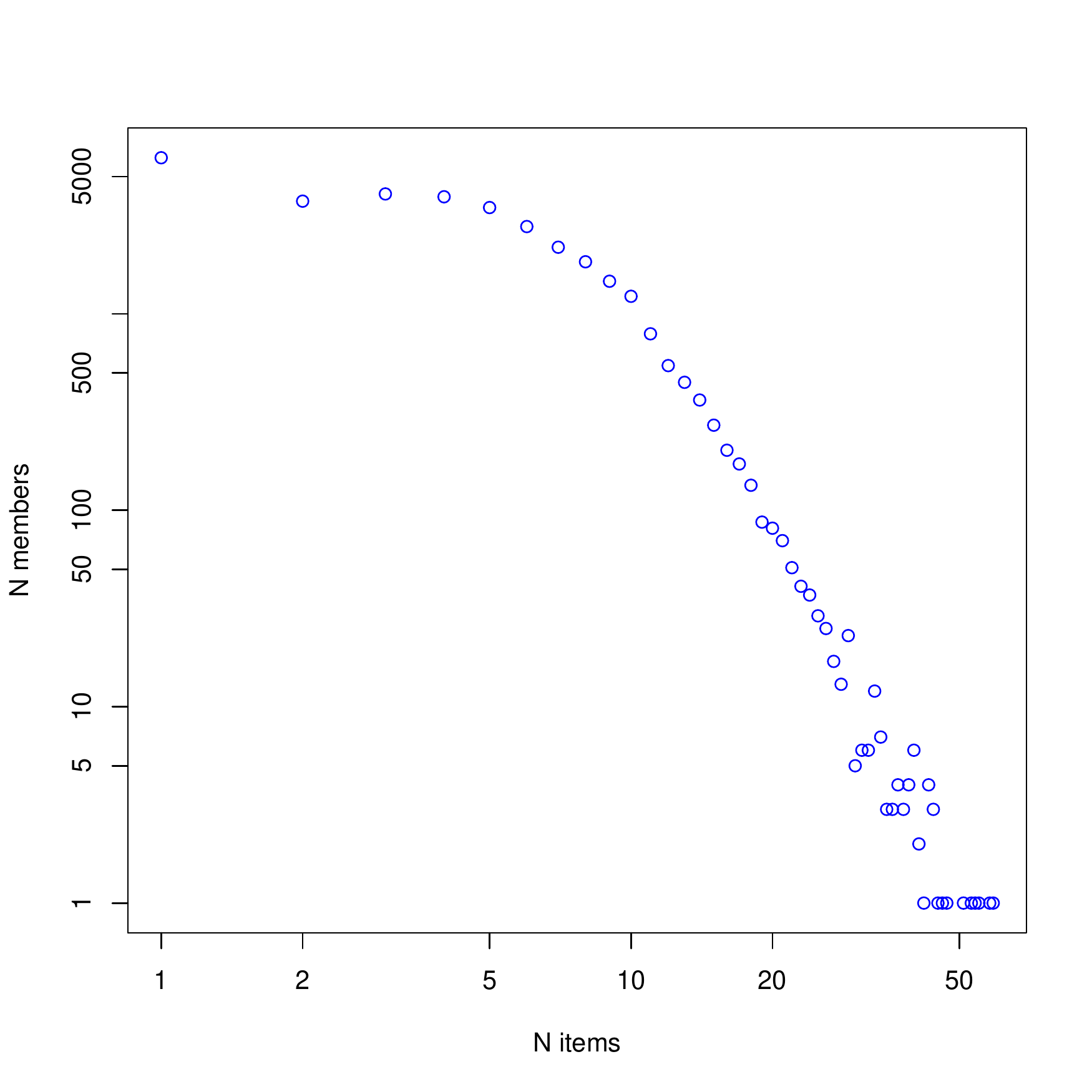}  
  \caption{Distribution of items per user}
  \label{fig:distribution}
\end{figure}

The proposed approach is studied on real world data extracted from the Viadeo
professional social network. The recommendation setting is the one described
in Section \ref{sec:real-world-illustr}: users can attach skills to their
profile. Skills are recommended to the users in order to help them build more
accurate and complete profiles. In this context, items are skills. The data
set used for the analysis contains 34 448 users and 35 741 items. The average
number of items per user is 5.33. The distribution of items per user follows
roughly a power law, as shown on Figure \ref{fig:distribution}. 

Both probabilities $P_t(u)$ and $P_t(i|u)$ are uniform. The quality function $l$ is
given by $l(g_t(u_{-i}),i)=\mathds{1}_{i\in g_t(u_{-i})}$ where $g_t(u_{-i})$
consists in 5 items. We use constant recommendation algorithms to focus on the
direct effects of our weighting proposal, which means here that each
algorithm is based on a selection of 5 items that will be recommended to all
users. 

The quality of a recommendation algorithm, $L_t(g)$, is estimated via stochastic
sampling in order to simulate what could be done on a larger data set than the
one used for testing. We selected repeatedly 20 000 users (uniformly among the
34 448, including possible repetitions) and then one item per user (according
to $P_t(i|u)$ or $P_t(i|u,\omega)$). 

The analysis is conducted on a 201 days period, from day 300 to day 500. Day 0
corresponds to the launch date of the skill functionality. As noted in Section
\ref{sec:real-world-illustr} two recommendation campaigns were conducted by
Viadeo during this period at $t=330$ and $t=430$ respectively. 

\subsection{Bias in action}\label{sec:bias-action}
We first demonstrate the effect of the bias on two constant recommendation
algorithms. The first one $g^1$ is modeled after the actual recommendation algorithm
used by Viadeo in the following sense: it recommends the five most recommended
items from $t=320$ to $t=480$. The second algorithm $g^2$ takes the opposite
approach by recommending the five most frequent items at time $t=300$ among
the items that were never recommended from $t=320$ to $t=480$. In a sense,
$g^1$ agrees with Viadeo's recommendation algorithm, while $g^2$ disagrees. 

\begin{figure}[htb]
  \centering
\includegraphics[width=\linewidth]{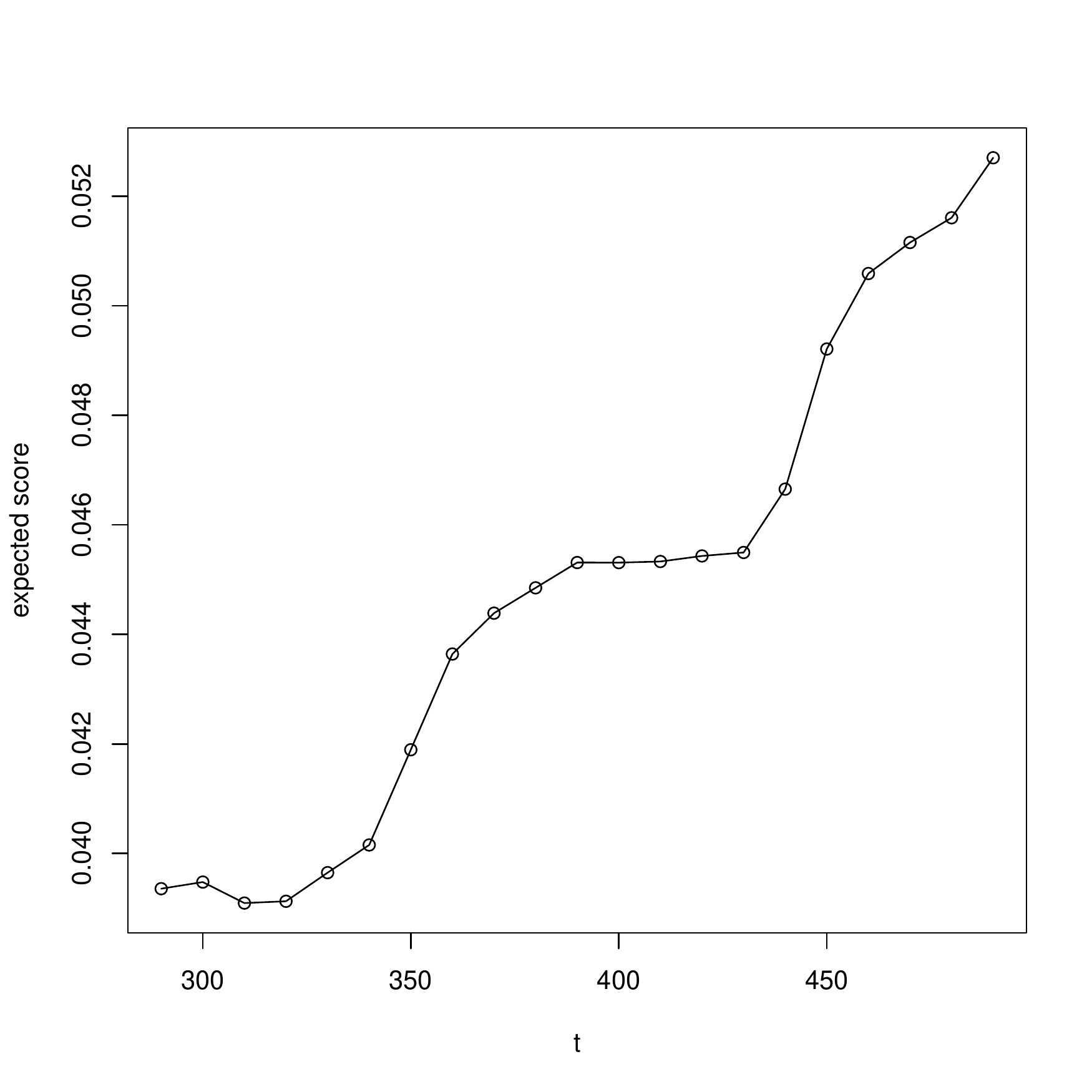}  
  \caption{Evolution of $L_t(g^1)$ over time ($g^1$ ``agrees'' with the
    recommendation algorithm)}
  \label{fig:g:one}
\end{figure}

\begin{figure}[htb]
  \centering
\includegraphics[width=\linewidth]{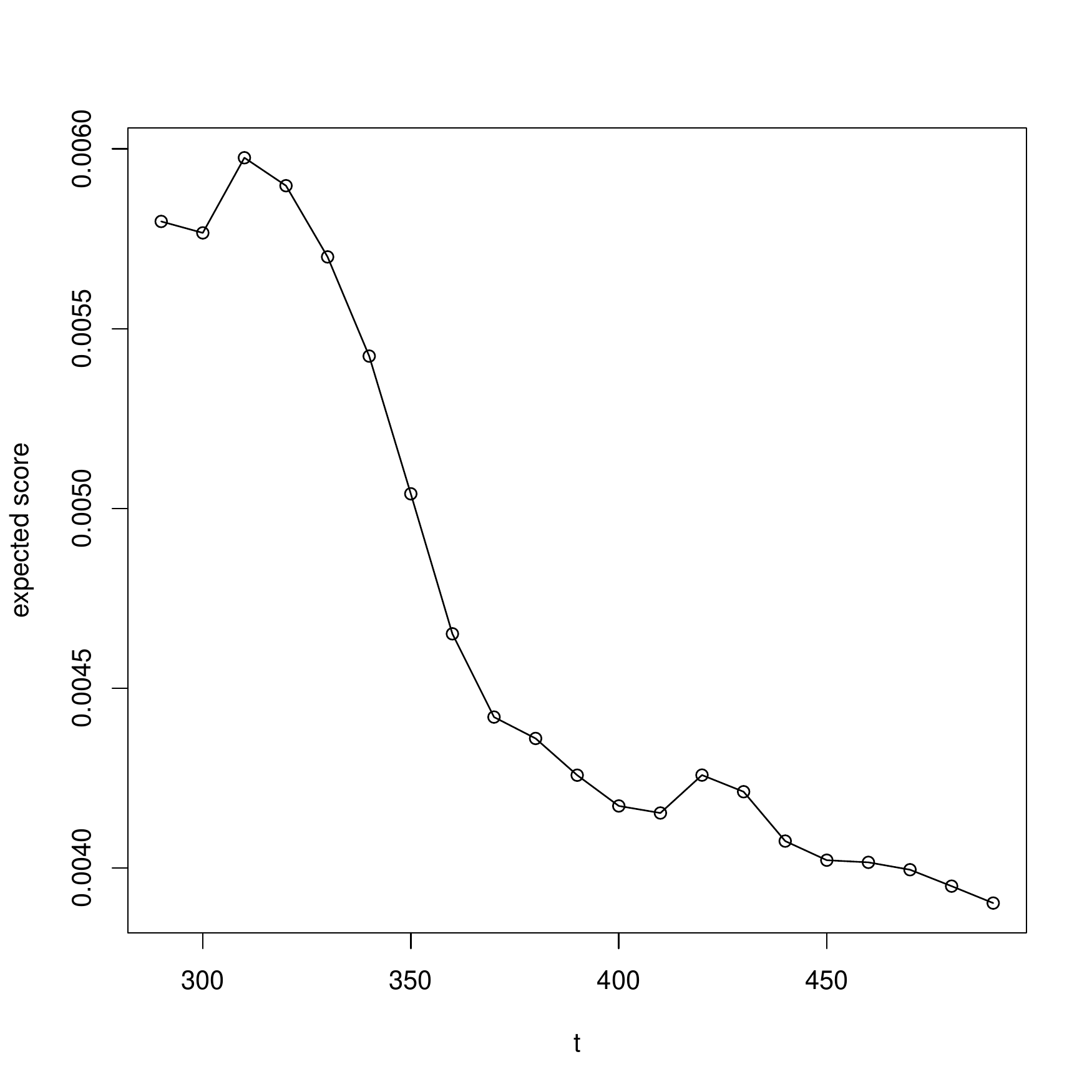}  
  \caption{Evolution of $L_t(g^2)$ over time ($g^2$ ``disagrees'' with the
    recommendation algorithm)}
  \label{fig:g:two}
\end{figure}

Figures \ref{fig:g:one} and \ref{fig:g:two} show the evolution of $L_t(g^1)$ and
$L_t(g^2)$ over time. As both algorithms are constant, it would be reasonable to
expect minimal variations of their offline evaluation scores. However in practice the estimated quality of $g^1$ increases by more than 25~\%, while the relative
decrease of $g^2$ reaches 33 \%.

\subsection{Reduction of the bias}
\begin{figure}[htb]
  \centering
\includegraphics[width=\linewidth]{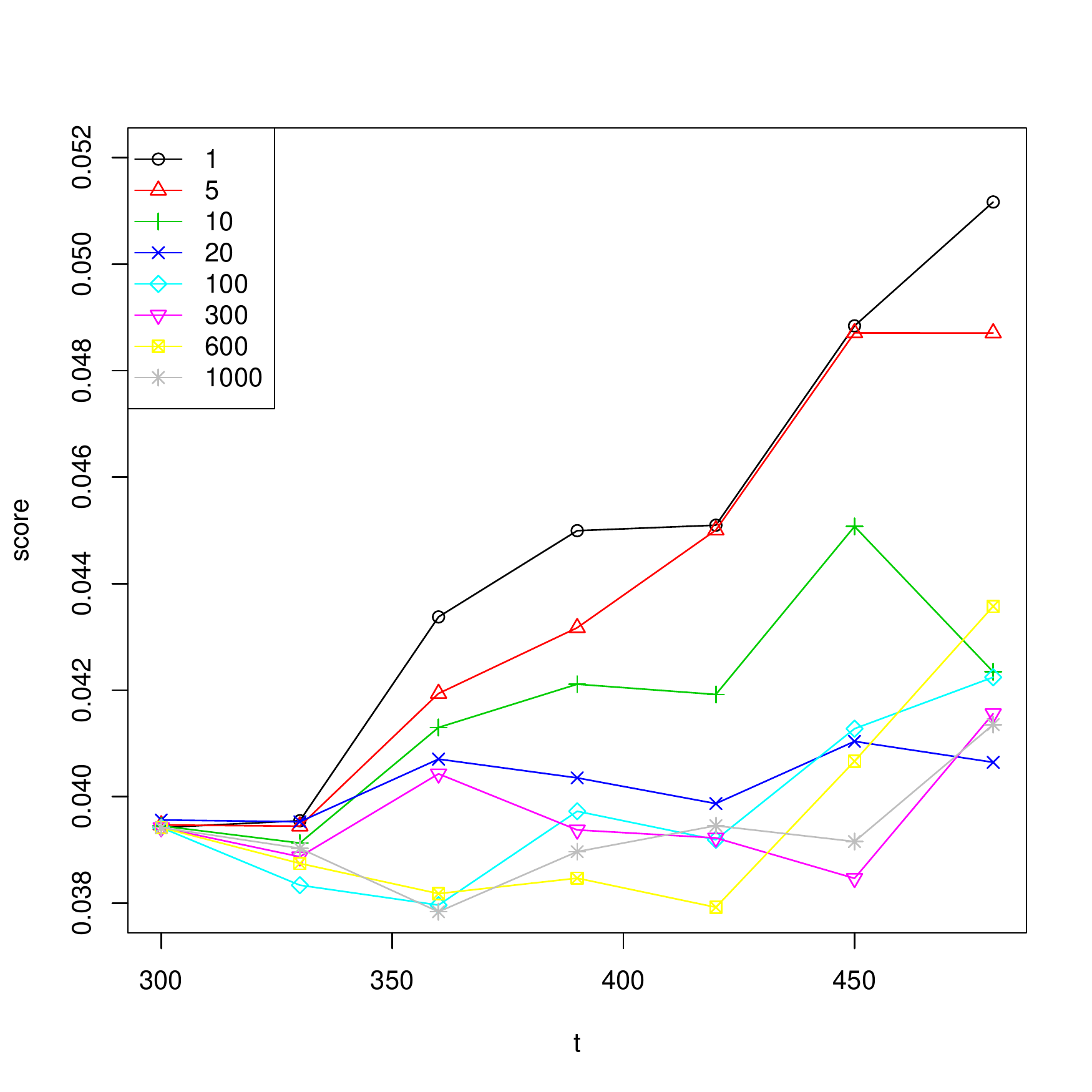}  
  \caption{Evolution of $L_t(g^1)$ over time ($g^1$ ``agrees'' with the
    recommendation algorithm) when items are weighted (see text for details).}
  \label{fig:g:one:weighted}
\end{figure}

\begin{figure}[htb]
  \centering
\includegraphics[width=\linewidth]{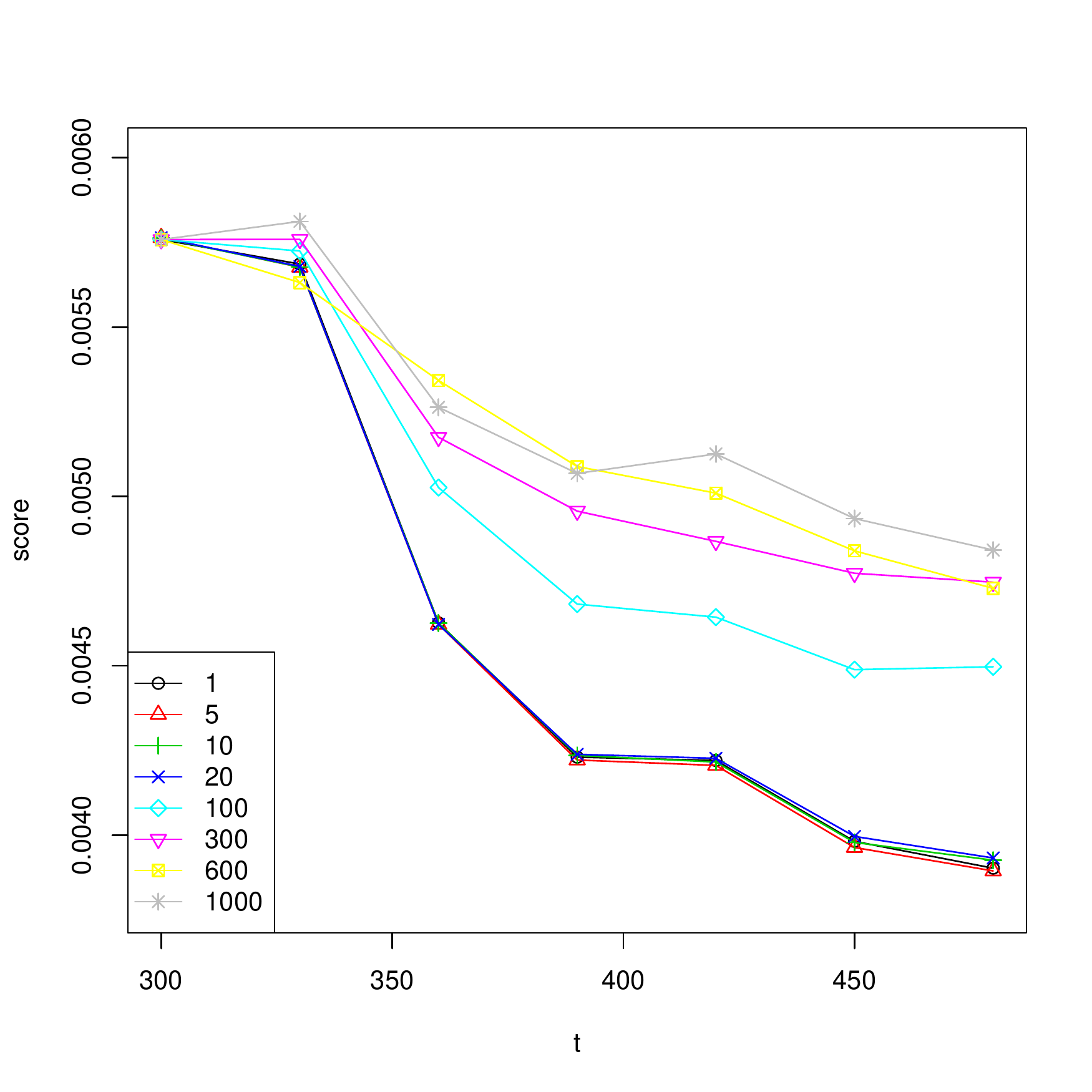}  
  \caption{Evolution of $L_t(g^2)$ over time ($g^2$ ``disagrees'' with the
    recommendation algorithm) when items are weighted (see text for details).}
  \label{fig:g:two:weighted}
\end{figure}

We apply the strategy described in Section~\ref{sec:reduc-eval-bias} to
compute optimal weights at different instants and for several values of the
$p$ parameter. Results are summarized in Figures \ref{fig:g:one:weighted} and
\ref{fig:g:two:weighted}. 

The figures show clearly the stabilizing effects of the weighting strategy on
the scores of both algorithms. In the case of algorithm $g^1$, the
stabilisation is quite satisfactory with only $p=20$ active weights. This is
expected because $g^1$ agrees with Viadeo's recommendation algorithm and 
therefore recommends items for which probabilities $P_t(i)$ change a lot
over time. Those probabilities are exactly the ones that are corrected by the
weighting technique. 

The case of algorithm $g^2$ is less favorable, as no stabilisation occurs with
$p\leq 20$. This can be explained by the relative stability over time of the
probabilities of the items recommended by $g^2$ (indeed, those items are not
recommended during the period under study). Then the perceived reduction in
quality over time is a consequence of increased probabilities associated to other
items. Because those items are never recommended by $g^2$, they correspond to direct
recommendation failures. In order to stabilize $g^2$ evaluation, we need to
take into account weaker modifications of probabilities, which can
only be done by increasing $p$. This is clearly shown by Figure
\ref{fig:g:two:weighted}. 

\section{Conclusion}
We have analyzed the offline evaluation bias induced by various factors that have influenced historical data as the recommendation algorithms previously used for such an online system. Indeed, as recommendations influence users, a 
recommendation algorithm in production tends to be favored by offline evaluation over time. On the contrary, an algorithm
with different recommendations will generally witness over time a reduction of its offline evaluation score. To overcome this bias, we have introduced a simple item weighting strategy inspired by techniques designed for tackling the
covariate shift problem. We have shown on real world data extracted from
Viadeo professional social network that the proposed technique reduces the evaluation bias for constant recommendation algorithms. 

While the proposed solution is very general, we have only focused on the
simplest situation of constant recommendations evaluated with a binary quality
metric (an item is either in the list of recommended items or not). Further works include the confirmation of bias reduction on more elaborate algorithms, possibly with more complex quality functions. The trade off between the computational cost of the proposed solution and its quality should also be
investigated in more details. 

\appendix
\section{Algorithmic details}
\subsection{Gradient calculation}
We optimize $D(\omega)$ with a gradient based algorithm and hence $\nabla D$
is needed. Let $i$ and $k$ be two distinct items $i\neq k$, then
\begin{align}
  \label{eq:nabla:Pi:k}\notag
\frac{\partial P(i|u,\omega)}{\partial
  \omega_k}&=-\frac{\omega_iP(i|u)P(k|u)}{\left(\sum_{j\in
      I}\omega_jP(j|u)\right)^2},\\
&=-P(i|u,\omega)\frac{P(k|u,\omega)}{\omega_k}.
\end{align}
We have also
\begin{equation}
  \label{eq:nabla:Pi:i}
\frac{\partial P(i|u,\omega)}{\partial \omega_i}=\frac{P(i|u,\omega)}{\omega_i}\left(1-P(i|u,\omega)\right),
\end{equation}
and therefore for all $k$:
\begin{equation}
  \label{eq:nabla:Pi:both}
  \frac{\partial P(i|u,\omega)}{\partial
  \omega_k}=\frac{P(k|u,\omega)}{\omega_k}\left(\delta_{ik}-P(i|u,\omega)\right).
\end{equation}
We have implicitly assumed that the evaluation is based on independent draws,
and therefore:
\begin{equation}
  \label{eq:notation:Pik}
P(i,k|\omega)=\sum_{u}P(i|u,\omega)P(k|u,\omega)P(u).
\end{equation}
Then
\begin{multline}
  \label{eq:nabla:D}\notag
\frac{\partial D(\omega)}{\partial \omega_k}=\\
\sum_{i}\frac{P_{t_0}(i)}{\omega_kP_{t_1}(i|\omega)}\left(P_{t_1}(i,k|\omega)-\delta_{ik}P_{t_1}(k|\omega)\right).
\end{multline}

Application: if $P(u) \sim \mathcal{U}(U)$ and $P(i|U) \sim \mathcal{U}(I_u)$, then:

\begin{eqnarray*}
 P(u) 	&=& \frac{1}{\#U}\\
 P(i|u) &=& \frac{1}{\#I_u} \cdot \mathds{1}_{i \in I_u}\\
 P(i|\omega) &=& \frac{1}{\#U} \cdot \sum_{u \in U_i }\frac{\omega_i}{\sum_{j \in I_u}\omega_j}\\
 P(i,k|\omega) &=& \frac{1}{\#U} \cdot \sum_{u \in U_i \cap U_k}\frac{\omega_i \omega_k}{(\sum_{j \in I_u}\omega_j)^2}
\end{eqnarray*}

Complexity: Assuming we have a sparse matrix $A \in \mathcal{M}{n_U, n_I}(\mathbb{R})$ such as $A_{u,i} = \mathds{1}_{u \in I_u}$, we suggest to precalculate $P_{t_0}(i)$ and then for each coordinate of the gradient and for each $i$:
\begin{itemize}
\item compute $P(i|\omega) = \sum_{u\in U_i} P(i|u,\omega)P(u)$ in $\mathcal{O}(\#U_i)$
\item compute $P(i,k|\omega) = \sum_{i\in U_i} P(i|u,\omega) P(k|u,\omega)P(u)$ in $\mathcal{O}(\#U_i)$
\end{itemize}

Then each $\frac{\partial D(\omega)}{\partial \omega_k}$ consists in a sum of $I$ terms computed in $\mathcal{O}(1)$, so that we can compute each coordinate of the gradient is $\mathcal{O}(\sum_{i \in I}U_i) = \mathcal{O}(|A|)$.

Thus, as $|A| = N_{U\times I}$ the complexity to compute $p$ coordinates of the gradient is $ \mathcal{O}(p\cdot N_{U \times I})$.

\bibliography{offline-evaluation-bias-bib}
\bibliographystyle{mlapa}

\end{document}